\documentclass[iop]{emulateapj}

\shorttitle{Absence of Cold Dust around HD~15407A}
\shortauthors{Fujiwara et al.}

\begin{document}

\title{The Absence of Cold Dust around Warm Debris Disk Star HD~15407A}


\author{
Hideaki~Fujiwara\altaffilmark{1}, 
Takashi~Onaka\altaffilmark{2}, 
Satoshi~Takita\altaffilmark{3}, 
Takuya~Yamashita\altaffilmark{4}, 
Misato~Fukagawa\altaffilmark{5}, 
Daisuke~Ishihara\altaffilmark{6}, 
Hirokazu~Kataza\altaffilmark{3}, 
and 
Hiroshi~Murakami\altaffilmark{3} 
}


\altaffiltext{1}{Subaru Telescope, National Astronomical Observatory of Japan, 650 North A'ohoku Pl., Hilo, HI 96720, USA;  hideaki@naoj.org}
\altaffiltext{2}{Department of Astronomy, School of Science, University of Tokyo, Bunkyo-ku, Tokyo 113-0033, Japan}
\altaffiltext{3}{Institute of Space and Astronautical Science, Japan Aerospace Exploration Agency, 
3-1-1 Yoshinodai, Chuo-ku, Sagamihara, Kanagawa 252-5210, Japan}
\altaffiltext{4}{National Astronomical Observatory of Japan, 2-21-1 Osawa, Mitaka, Tokyo 181-0015, Japan}
\altaffiltext{5}{Graduate School of Science, Osaka University, 1-1 Machikaneyama, Toyonaka, Osaka 560-0043, Japan}
\altaffiltext{6}{Department of Physics, School of Science, Nagoya University, Furo-cho, Chikusa-ku, Nagoya, Aichi 464-8602, Japan}


\begin{abstract}
We report {\it Herschel} and {\it AKARI} photometric observations at far-infrared (FIR) wavelengths of the debris disk around the F3V star HD 15407A, 
in which the presence of an extremely large amount of warm dust ($\sim 500$--600~K) has been suggested 
by mid-infrared (MIR) photometry and spectroscopy. 
The observed flux densities of the debris disk at 60--$160~\micron$ are clearly above the photospheric level of the star, 
suggesting excess emission at FIR as well as at MIR wavelengths previously reported. 
The observed FIR excess emission is consistent with the continuum level extrapolated from the MIR excess, 
suggesting that it originates in the inner warm debris dust and cold dust ($\sim 50$--$130$~K) is absent in the outer region of the disk.  
The absence of cold dust does not support a late heavy bombardment-like event 
as an origin of the large amount of warm debris dust around HD~15047A.
\end{abstract}




\section{Introduction}

Debris disks are considered as dusty disks replenished by the formation of the second generation dust 
during the main-sequence star phase, 
not as direct leftovers of protoplanetary disks \citep[e.g.][]{backman93}.  
Therefore debris disks hold hints of circumstellar rocky materials such as planets or their building blocks. 
Recent mid-infrared (MIR) observations by {\it Spitzer} and {\it AKARI} of high sensitivity 
have revealed the presence of warm debris disks that show MIR excess emission over the photosphere \citep[e.g.][]{fujiwara10}. 
Some warm debris disks possess a large amount of dust and cannot be explained 
by steady-state evolution of the disks by planetesimal collisions. 
They may be formed by transient events \citep{wyatt07}.
As transient events that generate bright warm debris disks, two major mechanisms are suggested. 
One is akin to the late heavy bombardment (LHB), which is suggested to account for the debris disk around $\eta$~Corvi \citep{lisse12}. 
The other is a two-body impact, which is akin to the giant impact that created the Moon, 
and suggested to account for the debris disk around BD~+20~307 \citep{weinberger11}. 
Examination of transient events relevant to debris disks is important not only 
for the understanding of dynamical evolution and presence of small bodies in the planetary system, 
but also for the understanding of the history of our solar system.

HD~15407A is an F3V main-sequence star at the distance of 55~pc from the Sun \citep{vanleeuwen07}, 
which possesses one of the most extreme warm debris disks.  
Large excess over the expected stellar photospheric emission at $\lambda \gtrsim 5~\micron$ 
has been reported by {\it IRAS}, {\it AKARI}, and {\it Spitzer} observations \citep{oudmaijer92,fujiwara12a}. 
The fractional luminosity (fraction of excess luminosity over stellar luminosity) of the MIR excess emission 
is estimated as $L_{\rm dust}/L_* \sim 0.005$. 
Fine dust of silica (SiO$_2$) and amorphous silicate are detected toward the star \citep{fujiwara12a}. 
The effective temperature and the metallicity of the star have been estimated 
as $6350$~K and [Fe/H] $= +0.08$, respectively, by Geneva-Copenhagen Survey \citep{holmberg09} 
and high-dispersion spectra obtained with Okayama Astrophysical Observatory/HIDES \citep{fujiwara12a}. 
As for the stellar age, two values, $80^{+40}_{-20}$~Myr \citep{melis10} 
and $2.1 \pm 0.3$~Gyr \citep{holmberg09}, have been suggested. 

In this Letter, we report new far-infrared (FIR) photometry of HD~15407A obtained with {\it Herschel} and {\it AKARI}, 
which constrains the amount of low-temperature dust around the star.   
We report the detection of FIR excess emission toward the star, which is accounted for 
solely by the MIR-emitting warm dust, suggesting the absence of cold dust around the star.  
We discuss the possible formation mechanism of the warm debris disk around HD~15407A.

\section{Observations and Data Reduction}

\subsection{{\it Herschel}/PACS Observations}

HD~15407A was observed on 2011 July 15 (UT) with Photodetector Array Camera and Spectrometer \citep[PACS;][]{poglitsch10}
onboard {\it Herschel} \citep{pilbratt10} 
using the photometer (scan map) mode (Observation ID: 1342224222--1342224225, PI: Ben Zuckerman)
with the three channels: $70~\micron$ (blue), $100~\micron$ (green), and $160~\micron$ (red).
%
We use the pipeline-processed Level-2 data products taken from the {\it Herschel} Science Archive.  
The star is seen clearly in the blue and green channel images and marginally in the red channel images. 
A very red source 
is also seen at the separation of $30\arcsec$ 
north of HD~15407A in the green and red images. 

Aperture photometry is conducted for all the individual images using IRAF {\verb apphot } task. 
The aperture radius and sky annulus dimensions are chosen as $5\arcsec$ and $50\arcsec$--$70\arcsec$, respectively, 
to obtain the best signal-to-noise ratios and the minimum contamination by the nearby source. 
Aperture correction is applied to the measured flux density using the aperture correction factors for a $5\arcsec$ aperture shown 
in \cite{pacsom11}. 
We calculate the weighted mean of the flux density of individual observations for each band. 
We take the $3\sigma$ of scatter of individual measurements as the photometric uncertainty for each band. 
The derived flux densities are shown in Table~\ref{photometry}. 
Color correction is not applied. The color-correction factors 
for sources of $\gtrsim 50$~K are 0.98--1.07 \citep{poglitsch10} and much less than the photometric errors. 

\subsection{{\it AKARI}/FIS Slow-scan Observations}

HD~15407A was observed on 2007 August 18 (UT) with the Far-Infrared Surveyor \citep[FIS; ][]{kawada07} 
onboard {\it AKARI} using the FIS01 observation mode as part of the director's time (Observation ID: 5110122-001). 
The FIS was operated in the photometry mode with the four bands: {\it N60} ($65~\micron$), 
{\it WIDE-S} ($90~\micron$), {\it WIDE-L} ($140~\micron$), and {\it N160} ($160~\micron$). 
The FIS data were processed with the FAST reduction toolkit for FIS slow-scan observation data, 
which is distributed within the {\it AKARI} internal team at present. 
Basic reduction procedure is the same as the FIS Slow-Scan Toolkit Version 20070914 officially provided. 
Among the four-band images, a source is seen at the position of HD~15407A 
only in the {\it N60} and {\it WIDE-S} images. 
The detected source is significantly affected by the dead pixel in the {\it N60} image 
and reliable photometry of the source is difficult for this band. 

Aperture photometry is conducted only for the {\it WIDE-S} images using IRAF {\verb apphot } task. 
The aperture radius and sky annulus dimensions are chosen as $40\arcsec$ and $120\arcsec$--$200\arcsec$, respectively. 
Aperture correction and flux calibration for point source 
are applied to the measured flux density as described in \cite{shirahata09}. 
Due to the large beam size of {\it AKARI}/FIS ($30\arcsec$ for {\it WIDE-S}), 
the nearby source detected by {\it Herschel}/PACS contaminates the signal. 
Since {\it Herschel}/PACS observations suggest that the signal count of the nearby source is about half of HD~15407A at $\sim 100~\micron$, 
the measured flux density in the {\it WIDE-S} band is divided by 1.5 to estimate the flux density of HD~15407A. 
The final estimated flux density of HD~15407A in the {\it WIDE-S} band of {\it AKARI}/FIS is shown in Table~\ref{photometry}. 

\subsection{Data from Literature and Catalog: {\it WISE}, {\it AKARI}/IRC, {\it IRAS}, Subaru/COMICS, {\it Spitzer}/IRS, 2MASS}

In addition to the FIR photometry described above, we collect available infrared photometric data from literature and catalogs. 
Near-infrared (NIR) $JHK_{\rm S}$ photometry is taken from Two Micron All Sky Survey (2MASS) Point Source Catalog \citep{cutri03}. 
NIR to MIR magnitudes at 3.4 ({\it W1} band), 4.6 ({\it W2} band), 12 ({\it W3} band), and $22~\micron$ ({\it W4} band) 
are taken from {\it Wide-Field Infrared Survey Explorer} ({\it WISE}) All-Sky Data Release \citep{cutri12}. 
Broad-band MIR photometric data at 9 ({\it S9W} band) and $18~\micron$ ({\it L18W} band) are taken 
from the {\it AKARI}/IRC All-Sky Survey Point Source Catalog \citep{ishihara10}.
Narrowband MIR photometric data with higher spatial resolution at 8.8, 11.7, and $18.8~\micron$ are also taken 
with Subaru/COMICS \citep{fujiwara12b}. 
{\it IRAS} photometric data in the MIR and FIR are taken from Point Source Catalog 
with the upper limits at 60 and $100~\micron$ \citep{beichman88}. 
All the collected flux densities are shown in Table~\ref{photometry}.





\section{Results}

\subsection{Spectral Energy Distribution and FIR Excess -- Absence of Cold Dust}

\begin{figure*}
\epsscale{1.0}
\plotone{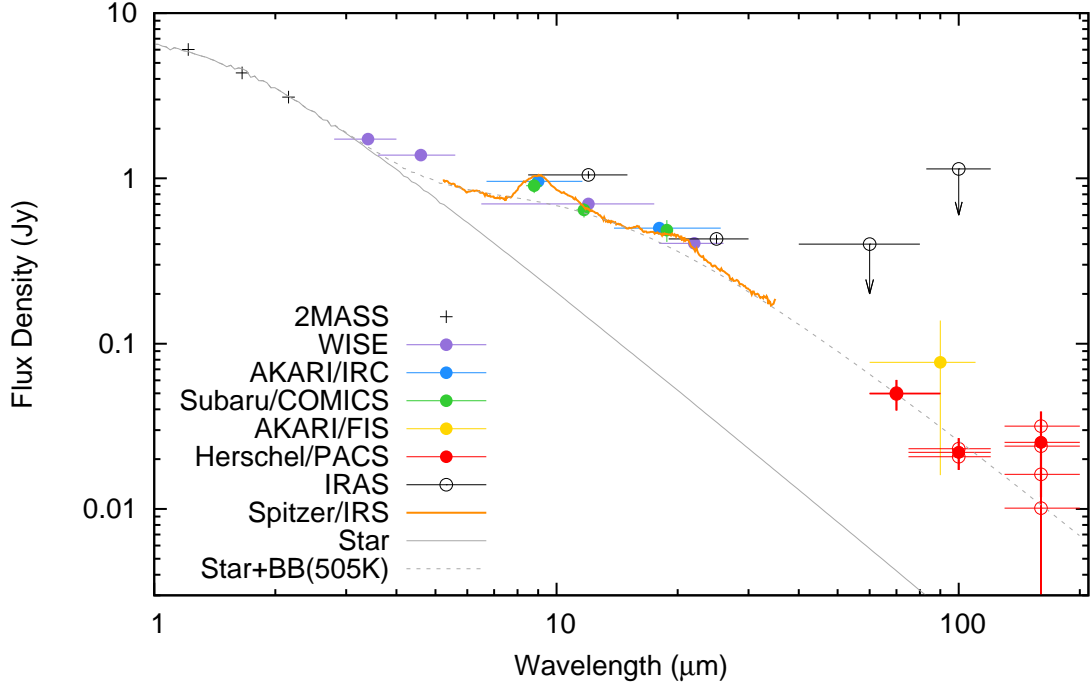}
\caption{NIR--FIR SED of HD~15407A. The black crosses indicate the 2MASS photometry, which determines the photospheric contribution of the star \citep{kurucz92} shown as the gray solid line. The purple, blue, green, yellow filled circles, and black open circles indicate the photometric data of {\it WISE}, {\it AKARI}/IRC, Subaru/COMICS, and {\it AKARI}/FIS, and {\it IRAS}, respectively. 
The red open and filled circles indicate the photometric measurements from individual images taken with {\it Herschel}/PACS 
and their weighted-average of photometry for each band, respectively. 
All the flux densities are not color-corrected. 
The vertical and horizontal bars of the photometric data show the flux density errors and the bandwidths, respectively. The orange solid and gray dashed lines indicate the {\it Spitzer}/IRS spectrum and the estimated continuum level of the excess emission, respectively, taken from \cite{fujiwara12a}. 
\label{SED_hd15407}}
\end{figure*}

The obtained spectral energy distribution (SED) of HD~15407A 
compiled with all the measured and collected photometric data in the NIR--FIR is shown in Figure~\ref{SED_hd15407}. 
We also plot the {\it Spitzer}/IRS spectrum at 5--$35~\micron$ and the photospheric emission 
of HD~15407A estimated from the Kurucz model \citep{kurucz92} 
with the effective temperature of $6500$~K and the surface gravity of $\log g=+4.0$ 
fitted to the 2MASS $K_{\rm S}$-band photometry \citep{fujiwara12a}. 
Significant excess emission over the photosphere in the FIR
is clearly shown by the {\it Herschel}/PACS and {\it AKARI}/FIS photometry, 
in addition to the large MIR excess previously reported.  
The detected flux densities at 70--$160~\micron$ are $\gtrsim 10$ times larger than the photospheric emission. 


In Figure~\ref{SED_hd15407}, we plot the continuum level of the excess emission 
(a blackbody of the temperature $T=505$~K with the fractional luminosity $L_{\rm dust}/L_* = 0.005$) 
estimated from the dust model fitted to the {\it Spitzer}/IRS spectrum at 5--$35~\micron$ \citep{fujiwara12a}. 
The {\it Herschel}/PACS flux densities at 70 and $100~\micron$ are consistent with 
the extrapolation of the estimated continuum level of the MIR excess 
within a $1\sigma$ photometric uncertainty, 
suggesting that the excess at 70 and $100~\micron$ originates in the inner warm debris dust. 
{\it AKARI}/FIS photometry at $90~\micron$ is also consistent 
with the estimated continuum level 
within a $1\sigma$ uncertainty. 
The consistency between the continuum estimated from the {\it Spitzer}/IRS spectrum (5--35$~\micron$) 
and the FIR photometry (70--100$~\micron$) suggests the absence of cold materials of $\sim 130$--50~K 
in the outer region ($R_{\rm dust}=10$--60~AU) of the HD~15407A system 
since such cold materials should show additional emission peaking at 35--100$~\micron$ over the continuum, 
which is not seen in the SED. 
A conservative upper limit on the fractional luminosity of cold dust around HD~15407A 
is set as $L_{\rm dust}/L_* \lesssim 2 \times 10^{-6}$, 
from a maximum thermal emission of additional blackbody of 50~K 
over the MIR excess emission allowed within the uncertainties of {\it Herschel}/PACS photometries. 
Although the $160~\micron$ photometry of {\it Herschel}/PACS lies slightly above the estimated continuum level, 
the uncertainty is large, and thus the presence of very cold dust ($\lesssim 30$~K) is not confirmed at present.

\subsection{Mass of Warm Debris around HD~15047A}

\cite{fujiwara12a} estimated the mass of fine warm dust around HD~15407A as $7 \times 10^{17}$~kg $\sim 10^{-7}M_\oplus$
from the dust model fitted to the band features seen in the {\it Spitzer}/IRS spectrum at 5--$35~\micron$. 
This value does not include 
the mass of the blackbody dust, which consists of large ($\gtrsim10~\micron$ in size) grains and rubbles.

Assuming that the continuum component of the observed excess emission at MIR and FIR wavelengths is attributable to blackbody grains of 
$505$~K at $R_{\rm dust} = 0.6$~AU as suggested by \cite{fujiwara12a} and this work 
and that its flux density is proportional to the sum of geometrical cross sections of the dust 
with the grain size distribution of $n(a) \propto a^{-3.5}$, 
we could estimate the total mass of large debris dust as 
\begin{eqnarray*}
\frac{M_{\rm total}}{M_\oplus} = 2.7 \times 10^{-4} \left( \frac{L_{\rm dust}}{L_*} \right) 
\left( \frac{\rho}{2.5~{\rm g~cm}^3} \right) \left( \frac{a_{\rm min}}{\micron} \right)^\frac{1}{2} \\
\times \left( \frac{a_{\rm max}}{\micron} \right)^\frac{1}{2} \left( \frac{R_{\rm dust}}{{\rm AU}} \right)^2, 
\end{eqnarray*}
\citep{hillenbrand08}, 
where $\rho$, $a_{\rm min}$, and $a_{\rm max}$ are the specific density and the minimum and maximum grain size of dust, respectively. 
Adopting $L_{\rm dust}/L_* = 0.005$ for the continuum excess emission from warm dust around HD~15407A, 
$\rho = 2.5$~g~cm$^{-3}$ of silicate, $a_{\rm min}=10~\micron$ as the smallest size of grain with featureless spectrum in the MIR, 
and $R_{\rm dust} = 0.6$~AU, the total mass is calculated as 
\begin{eqnarray*}
\frac{M_{\rm total}}{M_\oplus} = 1.5 \times 10^{-6} \left( \frac{a_{\rm max}}{\micron}\right)^\frac{1}{2}. 
\end{eqnarray*}
The maximum size $a_{\rm max}$ is not known though. The total mass of the warm debris dust around HD~15407A
is estimated as $\sim 5 \times 10^{-5} M_\oplus$ $\sim 3 \times 10^{20}$~kg assuming $a_{\rm max}=1$~mm. 
Assuming $a_{\rm max}=1$~km, the total mass is estimated as $\sim 0.05 M_\oplus$ $\sim 3 \times 10^{23}$~kg, 
corresponding to a few lunar masses or about a half of Mars mass.

An upper limit on the mass of cold dust ($\sim 50$~K at $R_{\rm dust} = 60$~AU) around HD~15047A could also be 
estimated in the same manner from the upper limit on the fractional luminosity $L_{\rm dust}/L_* \lesssim 2 \times 10^{-6}$
set by the present work. 
Assuming that the same grain size distribution ($n(a)$, $a_{\rm min}$, $a_{\rm max}$) as that of warm dust, 
an upper limit on the mass of cold dust is given as just a few times of the mass of warm dust as estimated above.

\subsection{NIR Excess}

\begin{figure*}
\epsscale{0.8}
\plotone{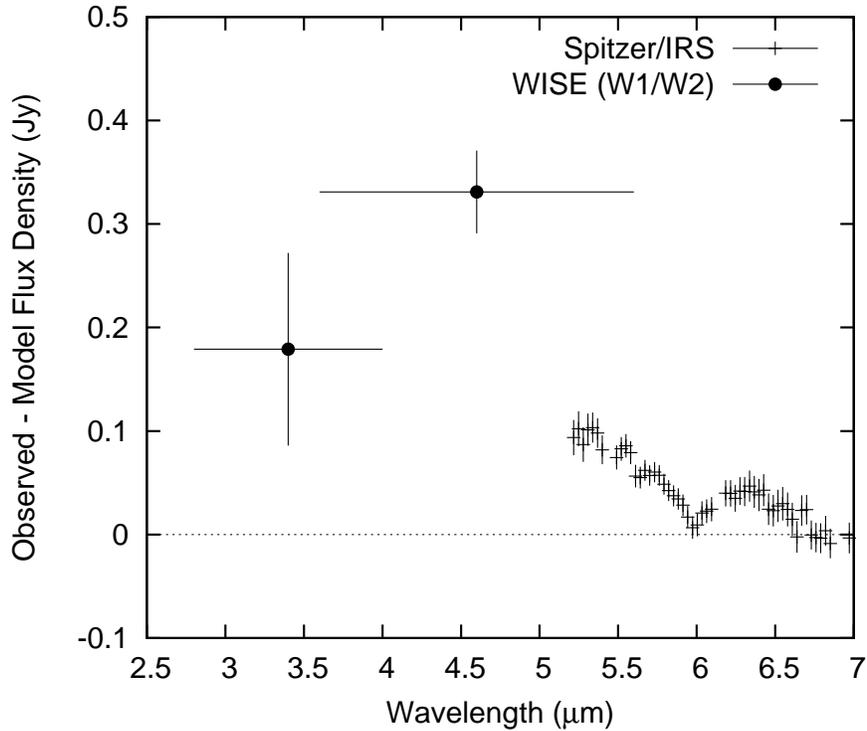}
\caption{NIR--MIR residual flux densities of HD~15407A subtracted 
by the model fitted to the {\it Spitzer}/IRS spectrum at 5--$35~\micron$. 
The crosses and filled circles indicate data from {\it Spitzer}/IRS and {\it WISE}, respectively. 
The vertical and horizontal bars of the photometric data show the flux density errors and the bandwidths, respectively. 
\label{NIR_SED_hd15407}}
\end{figure*}

As mentioned in the previous subsection, no debris material colder than $\sim 500$~K is suggested 
by the MIR-FIR SED of HD~15407A. 
On the other hand, photometry at $4.6~\micron$ in the {\it WISE} {\it W2}-band shows excess 
over the estimated continuum component of blackbody (Figure~\ref{SED_hd15407}), 
suggesting the presence of another component in addition to the warm dust components examined by \cite{fujiwara12a}. 
{\it Spitzer}/IRS observations of HD~15407A show that the residual spectrum subtracted 
by the best-fit model of the 5--$35~\micron$ excess increases toward $\lambda \lesssim 5~\micron$.
Weak excess emission at $3.4~\micron$ over the continuum component is also seen 
in the {\it WISE} {\it W1}-band data. 
Figure~\ref{NIR_SED_hd15407} plots the residual flux densities of excess emission at 3--$7~\micron$ 
subtracted by the warm (500--600~K) dust components derived in \cite{fujiwara12a}.
Although the residual flux densities seen around 3--$6~\micron$ might be attributable to 
hotter dust ($\gtrsim 1000$~K) in the vicinity of the star, 
the rise of the residual toward $\sim 4.6~\micron$ from its both sides seems too steep, 
suggesting the possible presence of a band feature attributable to circumstellar gas. 
A possible carrier of the NIR excess is the first overtone mode of SiO gas around $4~\micron$, 
which might be related to collisional event of rocky bodies (see Section~\ref{discussion}).

\section{Discussion}
\label{discussion}

MIR observations of HD~15407A by {\it AKARI}/IRC and {\it Spitzer}/IRS revealed 
the presence of a large amount of warm dust ($500$--$600$~K) 
at the distance of $\lesssim 1$~AU from the central star. 
FIR observations of the star by {\it Herschel}/PACS and {\it AKARI}/FIS in this work suggest 
remarkable results -- cold dust at $\gtrsim 10$~AU is absent 
and all of the measured 60--$100~\micron$ emission comes from the same population of dust grains 
that produce the bright MIR emission seen by {\it AKARI}/IRC and {\it Spitzer}/IRS. 
The results are contrary to a previous prediction by \cite{olofsson12} 
who suggested the presence of a population of cool dust around the star based on an MIR spectral model.
NIR--MIR observations by {\it WISE} also suggest the possible presence of hotter dust or gas component. 
In summary, the HD~15407A system possesses abundant warm dust at $\lesssim 1$~AU 
and no cold dust at $\gtrsim 10$~AU, which suggests that the system is quite different from 
most of the known debris disks with cold dust mainly discovered by {\it IRAS} \citep{rhee07}. 
So far about 10 bright warm debris disks around solar-type stars are known. 
Among those, HD~15407A seems similar to the G-star BD~+20~307, 
which possesses dust within 1~AU and no cold dust in the further region \citep{weinberger11}. 

An interesting issue of the debris disk around HD~15407A is its large fractional luminosity, 
which is estimated as $\sim 0.005$ solely from MIR observations by \cite{fujiwara12a}. 
This value is still secure even if the flux densities in the FIR are taken into account. 
Although the estimated age ranges from 80~Myr to 2~Gyr, 
the fractional luminosity of the disk is exceptionally ($\times 10^4$--$10^5$) larger 
than those predicted by steady-state models of planetesimal collisions for the suggested age range
\citep[$10^{-6}$--$10^{-8}$;][]{wyatt07}. 
Transient events are thus suggested to be responsible for the large amount of debris dust around HD~15407A. 

As the formation mechanism of warm debris disk around the $\sim 1$~Gyr old main-sequence star $\eta$~Corvi, 
an LHB-like event is suggested by \cite{lisse12}. 
Submillimeter observations by James Clerk Maxwell Telescope detect cold debris belt at $\sim 150$~AU around the star \citep{wyatt05}, 
which might form from planetesimals in the outer region (analog of Kuipe Belt objects). 
The MIR spectrum of the star suggests the presence of a large amount of warm debris dust, 
which might be produced at a few AU through collision(s) of an icy Kuiper Belt body or bodies 
falling from the outer region of the system \citep{lisse12}. 
In the scheme of an LHB event, the presence of a large amount of planetesimals in the outer region 
is required as impactors in the inner region. 
Self-grinding of the planetesimals should produce abundant small dust grains with low temperatures, 
which produce thermal emission at FIR wavelengths, as seen toward $\eta$~Corvi. 
$N$-body simulations of the LHB in the solar system by \cite{booth09} based on the Nice model \citep{gomes05} suggest 
that the mass surface density of debris at $\sim 20$--30~AU from the Sun 
would be $\sim10$ times larger than that at a few AU during the LHB. 
Therefore, the absence of cold dust around HD~15047A 
does not support an LHB-like event as a source of the warm dust around HD~15407A. 

Catastrophic collisions of two rocky, planetary-scale bodies in the terrestrial zone, 
which are an analog of the giant impact in the solar system, is 
suggested as a most likely source of warm debris around BD~+20~307 \citep{weinberger11}.
This giant-impact-like event might be applicable to the origin of the disk with warm debris 
and no cold material around HD~15407A, since a giant-impact-like event does not require 
a reservoir of planetesimals in the outer region. 
\cite{lisse09} conclude that the detection of silica dust and SiO gas features toward HD~172555 suggests 
a hypervelocity impact as the origin of the debris disk. 
Detection of abundant silica dust around HD~15407A would be harmonic with the possible mechanism of debris akin to that of HD~172555, 
but the presence of SiO gas around HD~15407A is not confirmed in its {\it Spitzer}/IRS spectrum \citep{fujiwara12a}.
Spectroscopic observations at 3--$5~\micron$ are needed to search for a hint of the first overtone feature of SiO gas. 

A giant impact is a probable hypothesis for the Moon formation around the Earth \citep{canup04} 
and is predicted to have been common during the final stage of terrestrial planet formation 
in extra-solar systems \citep{kenyon06}. A large amount of debris should be ejected 
in a giant impact. A recent theoretical study of the evocation of debris created in the Moon-forming giant 
impact \citep{jackson12} suggests that a giant impact generates a debris ring 
around the Earth orbit and that the fractional luminosity of the debris ring depends 
on the size of the largest fragment as well as the time after the impact. 
According to \cite{jackson12}, a fractional luminosity larger than 0.005 would be achieved until $100$~yr 
after the impact when the size of the largest fragment is 1--$10^3$~m 
and would not be achieved when the size of largest fragment is $> 10$~km. 
Assuming 100~yr as the lifetime of a giant-impact-generated bright debris disk , 
the probability to detect such an event is $\sim 10^{-6}$ and $10^{-7}$ 
for a star with age of 100~Myr and 1~Gyr, respectively. 
Even if we consider that the formation of an Earth-like planet requires around 10 giant impacts 
\citep[e.g.][]{kenyon06}, the probability increases only by one order of magnitude. 
HD 15407A is found as a possible giant-impact star among $\sim 600$ FGK dwarf stars 
in the {\it AKARI} survey \citep{fujiwara12b} and the apparent probability 
is $\sim 10^{-3}$, which is much larger than the value estimated above. 
The discrepancy might be due to an observational bias 
since MIR-luminous sources are easy to be found from the survey. A complete census of debris disks 
in a larger volume space by {\it WISE} may fill the gap. 


It should be noted that HD~15407A is in a possible binary system with the K2V star HD~15407B with the separation of $21\farcs2$ 
(projected distance of $1170$~AU) 
and HD~15407B might disturb the circumstellar material around HD~15407A dynamically. 
The study of the stability zone in a binary system by \cite{holman99} suggests that materials at $< 400$ and $< 150$ AU from HD~15407A are dynamically stable 
when the eccentricity of the system is 0.0 and 0.5, respectively. 
Thus the absence of cold materials at $\sim 10$--60~AU around the star is not due to the dynamical effect in the binary system.

\acknowledgments

This research is based on observations with {\it Herschel}, {\it AKARI}, {\it Spitzer}, {\it WISE}, {\it IRAS}, 2MASS, and Subaru Telescope. 
We thank C.M.\ Lisse, M.\ Shirahata, and the anonymous referee for their useful comments. 
This work was supported by KAKENHI (23103002).

{\it Facilities:} \facility{{\it Herschel} (ESA)}, \facility{{\it AKARI} (ISAS/JAXA)}, \facility{{\it Spitzer} (NASA)}, \facility{{\it WISE} (NASA)}, 
\facility{Subaru (NAOJ)}

\clearpage


\begin{table*}
\begin{small}
\begin{center}
\caption{Infrared Photometry of HD~15407A \label{photometry}}
\begin{tabular}{lrrr}
\tableline\tableline
                    & Wavelength  & Flux Density & Photosphere\\
Instrument (Filter) & ($\micron$) & (Jy)        & (Jy)                          \\
\tableline
2MASS ($J$)         & 1.22        & $6.017 \pm 0.181 $ & \nodata                      \\
2MASS ($H$)         & 1.65        & $4.352 \pm 0.136 $ & \nodata                      \\
2MASS ($K_{\rm S}$) & 2.16        & $3.105 \pm 0.062 $ & \nodata                      \\
{\it WISE} ({\it W1}) & 3.4       & $1.729 \pm 0.093 $ & 1.53                         \\
{\it WISE} ({\it W2}) & 4.6       & $1.381 \pm 0.040 $ & 0.88                         \\
{\it WISE} ({\it W3}) & 12        & $0.701 \pm 0.010 $ & 0.12                         \\
{\it WISE} ({\it W4}) & 22        & $0.405 \pm 0.009 $ & 0.04                         \\
{\it AKARI}/IRC ({\it S9W})  & 9  & $0.960 \pm 0.030 $ & 0.22                         \\
{\it AKARI}/IRC ({\it L18W}) & 18 & $0.500 \pm 0.020 $ & 0.06                         \\
Subaru/COMICS ({\it N8.8})  & 8.8  & $0.904 \pm 0.090 $ & 0.26                        \\
Subaru/COMICS ({\it N11.7}) & 11.7 & $0.644 \pm 0.064 $ & 0.15                        \\
Subaru/COMICS ({\it Q18.8}) & 18.8 & $0.486 \pm 0.073 $ & 0.06                        \\
{\it IRAS}                  & 12  & $1.050 \pm 0.060 $ & 0.12                         \\
{\it IRAS}                  & 25  & $0.430 \pm 0.030 $ & 0.03                         \\
{\it IRAS}                  & 60  & $<0.400$             & 0.005                        \\
{\it IRAS}                  & 100 & $<1.140$            & 0.002                        \\
{\it AKARI}/FIS ({\it WIDE-S}) & 90 & $0.077 \pm 0.061 $ & 0.003                      \\
{\it Herschel}/PACS (Blue)  & 70  & $0.050 \pm 0.011 $ & 0.004                        \\
{\it Herschel}/PACS (Green) & 100 & $0.022 \pm 0.005 $ & 0.002                        \\
{\it Herschel}/PACS (Red)   & 160 & $0.025 \pm 0.014 $ & 0.0008                       \\
\tableline
\end{tabular}
\end{center}
\end{small}
\end{table*}

\clearpage


\clearpage

\end{document}